\documentclass[twocolumn,trackchange]{aastex631}

\usepackage{amsmath}
\usepackage{mathtools}
\usepackage[leqno,fleqn,intlimits]{empheq}
\usepackage{bm}
\usepackage{empheq}
\usepackage[inline]{enumitem}
\usepackage{cleveref}
\usepackage{tabularx}
\usepackage{svg}
\usepackage{threeparttable} 

\begin{document}

\title{The role of ammonia in the  distribution of volatiles in
the primordial hydrosphere of Europa}

\author{Alizée Amsler Moulanier}
\affiliation{Aix-Marseille Université, CNRS, CNES, Institut Origines, LAM, Marseille, France}

\author{Olivier Mousis}
\affiliation{Aix-Marseille Université, CNRS, CNES, Institut Origines, LAM, Marseille, France}
\affiliation{Institut Universitaire de France (IUF)}

\author{Alexis Bouquet}
\affiliation{Aix-Marseille Université, CNRS, Institut Origines, PIIM, Marseille, France}
\affiliation{Aix-Marseille Université, CNRS, CNES, Institut Origines, LAM, Marseille, France}

\author{Christopher R. Glein}
\affiliation{Space Science Division, Space Sector, Southwest Research Institute, 6220 Culebra Road, San
Antonio, TX 78238-5166, United States}



\begin{abstract}

The presence of a hydrosphere on Europa raises questions about its habitability, and studies of its volatile inventory can provide insight into its formation process. Different scenarios suggest that Europa's volatiles could be derived from cometary ices or devolatilized building blocks. The study of post-accretion processes, in particular the "open ocean" phase that likely occurred before the formation of the icy crust, is crucial to distinguish these origins, as this phase is likely to have influenced the volatile inventory. The abundance of ammonia in Europa's building blocks is also crucial for understanding the composition of its ocean and primordial atmosphere.
We aim to investigate ocean-atmosphere equilibrium during the post-accretion period by varying the ammonia fraction in the atmosphere. Our model evaluates the vapor-liquid equilibrium of water and volatiles, as well as the chemical equilibrium within the ocean, to study Europa's early hydrosphere. We explore two initial conditions: one in which Europa's hydrosphere originates from comet-like building blocks, and another in which it forms in equilibrium with a thick, CO$_2$--rich atmosphere. In both scenarios, the initial ratio of accreted CO$_2$ to NH$_3$ determines the magnitude of their partial pressures in Europa's early atmosphere. If this ratio exceeds a certain threshold (set to $10^{-4}$ in this study), the atmosphere will be CO$_2$--rich; otherwise, it will be CO$_2$--depleted by multiple orders of magnitude. Overall, our work provides a initial assessment of the distribution of primordial volatiles in Europa's primitive hydrosphere, and provides a baseline for interpreting data from the upcoming \textit{Europa Clipper} mission.

\end{abstract}


\section{Introduction} \label{sec:intro}

Jupiter's moon Europa harbors a vast global subsurface ocean beneath its icy outer shell \citep{khurana_induced_1998,pappalardo_does_1999,kivelson_galileo_2000}. This ocean is likely in contact with the moon's silicate mantle, potentially facilitating hydrothermal processes \citep{zolotov_chemical_2009}. Interactions between the liquid water and the surface of the icy  crust are anticipated, contributing to the exchange of endogenous materials. Indeed, Europa's average surface is young \citep{bierhaus_europas_2009} and the presence of salts on disrupted areas of Europa's surface could be indicative of such processes \citep{mccord_brines_2002,dalton_spectral_2005}. 

The circumstances surrounding Europa's formation remain a subject of ongoing interest, especially concerning the origin of its water content ($\sim$8 wt$\%$ \citep{schubert_interior_2004, gomez_casajus_updated_2021}), which could derive from different type of accreted material and source location. Europa's volatile content might have come directly from the circumplanetary disk (CPD), where its building blocks could have condensed in situ \citep{Prinn81}. Another possibility is that Europa's volatiles originated in the protosolar nebula (PSN), where its building blocks formed before entering the Jovian CPD and subsequently accreting onto the Galilean moons \citep{canup_formation_2002,mousis_constraints_2004,ronnet_saturns_2018}.Moreover, water could have been brought to Europa through different processes. An accretion of a mixture of hydrated rocks and ice is a plausible scenario modeled in \cite{canup_formation_2002} and \cite{ronnet_pebble_2017}. On the other hand, after carbonaceous chondrites dehydration the accretion process could also be at the origin of Europa's hydrosphere \citep{melwani_daswani_metamorphic_2021,trinh_slow_2023,mousis_early_2023}. 

Understanding the processes related to the early stages of Europa's evolution is crucial for deducing the moon's formation history based on its present composition. It is conceivable that shortly after its accretion, Europa existed in an ``open-ocean'' configuration, where a dense atmosphere coexisted and chemically equilibrated with the global ocean \citep{lunine_clathrate_1987,lunine_massive_1992}. In time, such a configuration would evolve to the ``closed-ocean'' one, observable nowadays, where the global ocean is covered by an ice crust. As the current composition of Europa's subsurface ocean may be intrinsically linked to the ``open-ocean'' one, it is crucial to retrace the evolution of the hydrosphere. The coupling of the modeling of how volatiles were distributed during the ``open-ocean'' phase, and the future detection of compounds indicative of the subsurface ocean's current composition, could provide valuable insights into the evolution of the volatile content possibly sequestered in Europa's contemporary hydrosphere (global ocean and ice crust). 
However, the volatile content incorporated into Europa's ocean after accretion, is contingent on the type of accreted material and thus the moon's formation conditions. 
As a consequence, through the exploration of this early ``open ocean'' phase in Europa's history we intend to quantify the consequences of different formation scenarios  on the evolution of the volatile content of Europa's hydrosphere and confront in the future those results to compositional data from ESA's \textit{JUICE} and NASA's \textit{Europa Clipper} missions \citep{pappalardo_science_2024}.

Observations of Europa's surface have led to suspicions of the presence of compounds such as CO$_2$, likely of endogenic origin \citep{trumbo_distribution_2023,villanueva_endogenous_2023}, sulfates \citep{hibbitts_color_2019}, and Mg-chlorinated salts \citep{ligier_vltsinfoni_2016} within the ocean. While not yet detected, it is important to assess the distribution of nitrogen-bearing species. These species can significantly influence habitability conditions, N being one of the essential elements to build life as we know it.
One relevant nitrogen-bearing species worth studying is ammonia, which is abundant in cometary ices \citep{bockelee-morvan_composition_2017}, consequently, it is presumed to be present among the nitrogen-bearing species found on icy worlds. NH$_3$ is also frequently cited as an antifreeze in the oceans of icy moons \citep{grasset_cooling_1996,neveu_aqueous_2017,trinh_slow_2023}, so it could play a role in maintaining their liquid state.

Moreover, NH$_3$ can influence the pressure distribution within the primordial atmosphere. As CO$_2$ and NH$_3$ react in water, their speciation products which are respectively acidic and basic will engage in acid-base reactions. These ionic forms are more soluble than their neutral counterparts, suggesting that the chemical equilibrium may shift based on whether CO$_2$ or NH$_3$ predominates. As a result, depletion of either CO$_2$ or NH$_3$ in the liquid phase would cause the partial pressure distribution in the gas phase to adjust accordingly. In addition, experimental evidence has shown a correlation between the dissolved CO$_2$--NH$_3$ ratio and the partial pressure distribution in the H$_2$O--CO$_2$--NH$_3$ system's liquid-vapor equilibrium \citep{van_krevelen_composition_1949,goppert_vaporliquid_1988,bieling_evolutionary_1989}. Previous studies, such as \cite{melwani_daswani_metamorphic_2021}, predicted the existence of a thick CO$_2$ primordial atmosphere for Europa. However, those models did not incorporate ammonia or nitrogen in the primordial bulk composition of the moon. We aim to explore how the presence of ammonia impacts the initial distribution of volatiles in the primordial hydrosphere. The amount and chemical forms of nitrogen delivered to Europa during its formation depend on the composition of the materials accreted throughout its growth phase. Nitrogen compounds, although present in different concentrations, have been detected in both carbonaceous chondrites and cometary ice \citep{marion_modeling_2012,pizzarello_ammonia_2012,rubin_elemental_2019}. In this study, we examine two scenarios for the composition of Europa's primordial atmosphere. Both scenarios assume that the atmospheric reservoir outgassed from the interior due to vigorous accretional and radiogenic heating, reaching equilibrium with an underlying liquid ocean \citep{lunine_clathrate_1987,bierson_explaining_2020}. The first scenario assumes a comet-like atmospheric composition resulting from the outgassing of building blocks that originated in the PSN \citep{canup_formation_2002,mousis_constraints_2004,ronnet_pebble_2017}. The second scenario posits a primordial, CO$_2$-rich atmosphere formed after Europa’s accretion, as established by \cite{melwani_daswani_metamorphic_2021}.

To highlight the implications of such a choice on the initial distribution of volatiles in the early phase of the moon, we followed the approach of \cite{marounina_role_2018} by developing a model designed to calculate volatile partitioning between the ocean and the atmosphere. This model takes into account liquid-vapor equilibrium between the two phases, and is coupled with chemical equilibrium in the liquid phase. We employ this model to evaluate the extent to which the initial distribution of volatiles, particularly CO$_2$ and NH$_3$ content, influences the equilibrium between the primordial atmosphere and ocean. 

The paper proceeds as follows. Section \ref{sec:model} is devoted to a description of the model used in this study to compute the distribution of volatiles between an ocean and an atmosphere. In Section \ref{sec:results} we first show the proportion of volatiles in early Europa's hydrosphere, for the specific case where Europa's water content was brought by cometary ice, and explore how the amount of incorporated ammonia may shift this equilibrium. Then, the equilibrium that would result starting from a primordial CO$_2$-rich atmosphere is computed as well. In Section \ref{sec:discussion}, we discuss the implications of our results in the context of the current state of knowledge about Europa. Finally, our results are summarized in Section \ref{sec:summary} and their outcomes along with additional processes to be considered are discussed.

\section{Model description} \label{sec:model}

\begin{figure*}
\centering
\includegraphics[width=0.9\linewidth]{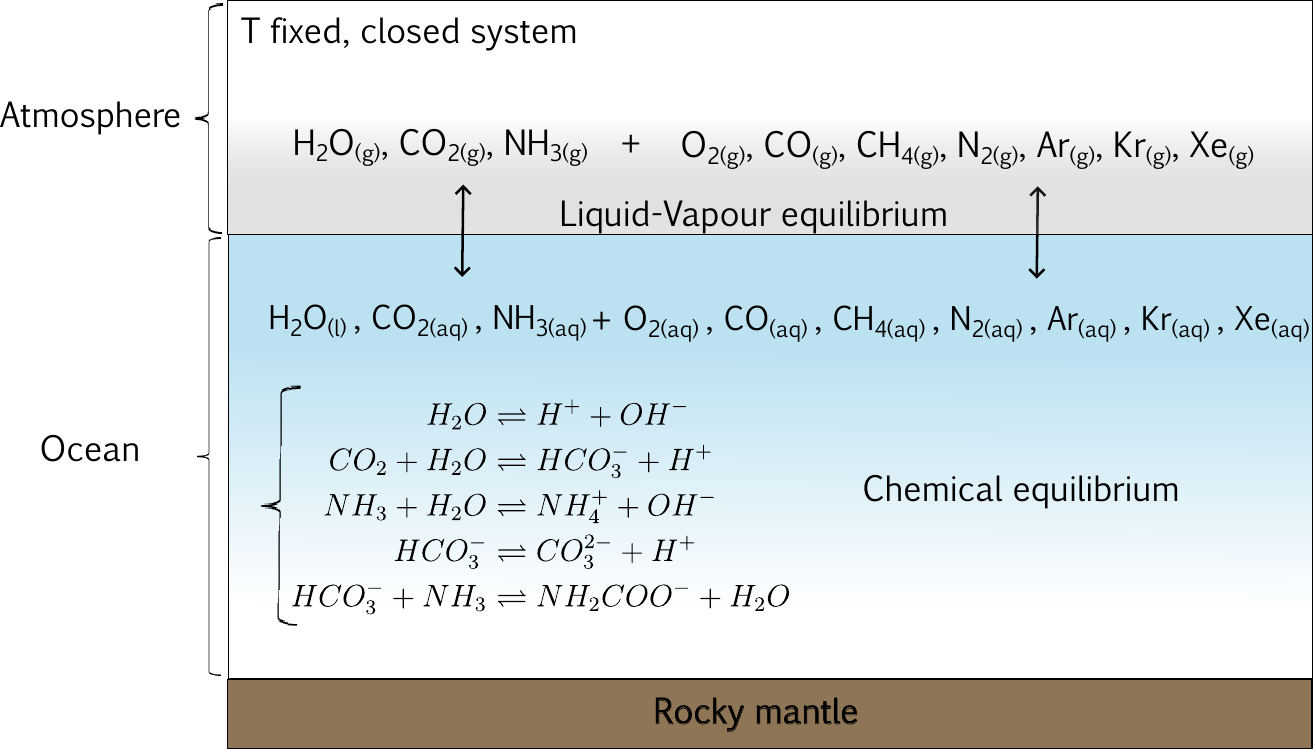}
\caption{Model scheme representation. The model represents the dissolution of volatiles from the primordial atmosphere into the ocean. The liquid-vapor equilibrium of the volatile species is coupled to a chemical equilibrium model that considers the speciation of H$_2$O, CO$_2$, and NH$_3$. This coupled approach allows the partitioning of the volatiles between the atmosphere and the ocean to be determined, taking into account both physical and chemical equilibrium. This model does not account for water-rock interactions between the ocean and silicate mantle.}
\label{fig:LVmodel}
\end{figure*}

The model used in this study, summarized in Fig. \ref{fig:LVmodel}, focuses on the dissolution of volatiles in a water ocean at a shallow depth. Two thermodynamic phases -- the primordial atmosphere and the ocean -- are in equilibrium, and the dissolved gases are considered as aqueous species. Temperature is a user input and assumed to be constant at the ocean-atmosphere interface, within a range of temperatures possible at the surface after accretion \citep{schubert_internal_1981,bierson_explaining_2020}.  
The approach involves initiating the volatile species into the atmosphere, acting as a primordial reservoir in equilibrium with the ocean during its early open phase. This is justified by the fact that, immediately after accretion, the atmosphere formed through outgassing, releasing a substantial portion of the volatiles initially present in Europa’s building blocks into the atmosphere \citep{lunine_massive_1992,melwani_daswani_metamorphic_2021}. Atmospheric escape is not considered in this study. We focus on the interface between the atmosphere and the ocean.
The non-ideal behavior and interactions of volatiles in the atmosphere and the ocean are accounted for by computing fugacity and activity coefficients. The equilibrium between the two phases is coupled to a chemical equilibrium model, in which the speciation of the two most reactive species -- carbon dioxide and ammonia -- is considered. The modeled chemical system is simplified, in order to understand its general features of its chemistry. Species such as O$_2$, CO, CH$_4$, and N$_2$ are not taken into account due to either their low reactivity in water or because the conditions for chemical reactions to happen are not met in this modeling (for CO oxidation, for instance \citep{seewald_experimental_2006}). They are present however in the liquid-vapor equilibrium. We do not consider as well any water-rock reactions involving Europa's silicate mantle and ocean. The description of the model is split into two sub-sections, the first is focused on the liquid-vapor equilibrium and the second on the chemical speciation within the ocean. 

\subsection{Liquid-Vapor equilibrium}

We first introduce the equations for phase equilibrium. The liquid-vapor equilibrium for each species is set by the equality between the vapor and liquid phases' fugacities, $f_i^V$ and $f_i^L$, respectively:

\begin{equation}
f_i^V(T,P,y_i) = f_i^L(T,P,x_i).
\label{Eq:raoult}
\end{equation}

\noindent Here, the fugacity of the species $i$ in the vapor phase, $f_i^V$, is a function of the temperature $T$, pressure $P$, and mole fraction $y_i$ in the vapor phase. Similarly, the fugacity of the species $i$ in the liquid phase, $f_i^L$, is a function of $T$, $P$, and mole fraction $x_i$ in the liquid phase. 

This equation represents Raoult's law, which describes the equilibrium between the vapor and liquid phases for a given species. By enforcing this equality, the model can determine the partitioning of volatiles between the atmosphere and the ocean.

By introducing the fugacity coefficient $\phi$ and the activity coefficient $\gamma$, the fugacity of each phase for an individual species $i$ can be written as:

\begin{equation}
\begin{split}
f_i^V = \phi_iy_iP, \\
f_i^L = \gamma_ix_if_i^0
\end{split}
\label{Eq:fug}
\end{equation}

\noindent In the vapor phase, the fugacity $f_i^V$ is a function of the fugacity coefficient $\phi_i$, the mole fraction $y_i$, and the total pressure $P$ in Pascal. This expression accounts for the non-ideal behavior of the gas phase. In the liquid phase, the fugacity $f_i^L$ is a function of the activity coefficient $\gamma_i$, the mole fraction $x_i$, and the standard-state fugacity $f_i^0$. The activity coefficient $\gamma_i$ captures the non-ideal interactions between the dissolved species in the liquid phase.

By substituting these expressions into Eq. \eqref{Eq:raoult}, the model can determine the partitioning of volatiles between the atmosphere and the ocean, taking into account the non-ideal behavior of both phases. The standard-state fugacity $f_i^0$ of H$_2$O is set to the saturation pressure of water, $P_{H_2O}^s(T)$, which is calculated using the Antoine equation \citep{stull_vapor_1947}: 

\begin{equation}
    \log_{10}(P^{s}_{H_2O}) = A - \cfrac{B}{T+C},
    \label{Eq:antoine}
\end{equation}

\noindent with $A=4.6543$, $B=1435.264$, $C=-64.848$, \textit{T} in K and \textit{P} in bar.

Equation \eqref{Eq:fug} can then be rewritten as:
\begin{equation}
\begin{split}    
&\phi_{H_2O} \; P\; y_{H_2O} = \gamma_{H_2O}\; x_{H_2O}\; P_{H_2O}^s(T) \\
&\phi_{i} P y_{i} = \gamma_{i}\; x_{i}\; H_{H_2O,i}(T)\exp(\cfrac{v_{i,H_2O}^{\infty}(P-P_{H_2O}^s)}{RT}),
\end{split}
\label{Eq:eq_LV_equation}
\end{equation}

\noindent where the standard-state fugacity $f_i^0$ of dissolved species $i$ is approximated as the Henry's law constant, corrected by a Poynting factor \citep{kawazuishi_correlation_1987,darde_modeling_2010}. The Henry's law constants and their temperature dependencies are reported {in Table \ref{tab:henrys_value}} and were taken from \cite{kawazuishi_correlation_1987} for CO$_2$ and NH$_3$, and from \cite{warneck_atmospheric_2012} for the other species. 

\begin{table*}[htpb]
\centering
\begin{threeparttable}[b]

\caption{Henry's constants interpolation formulas of gases in water.}
\begin{tabular}{@{}ccc@{}}
\hline
\hline
\smallskip
Molecule & Henry's constant formula & Unit\\
\hline
\smallskip
CO$_2$\tnote{a}   & ln(H$_i$)= $-17060.71/T - 68.31596\ln{T} + 0.06598907T + 430.1920$   & bar.kg.mol$^{-1}$      \\
NH$_3$\tnote{a}   & ln(H$_i$)= $-7579.948/T - 13.58857\ln{T} + 0.008596972T + 96.23184$   & bar.kg.mol$^{-1}$  \\ 
O$_2$\tnote{b}    & ln(H$_i$)= $-175.33 + 8747.5/T + 24.453\ln{T}$  & mol.dm$^{-3}$.atm$^{-1}$     \\ 
CO\tnote{b}       & ln(H$_i$)= $-178.00 + 8750.0/T + 24.875\ln{T}$  & mol.dm$^{-3}$.atm$^{-1}$      \\ 
CH$_4$\tnote{b}   & ln(H$_i$)= $-211.28 + 10447.9/T + 29.780\ln{T}$ & mol.dm$^{-3}$.atm$^{-1}$      \\ 
N$_2$\tnote{b}    & ln(H$_i$)= $-177.57 + 8632.1/T + 24.798\ln{T}$  & mol.dm$^{-3}$.atm$^{-1}$      \\ 
Ar\tnote{b}       & ln(H$_i$)= $-146.40 + 7479.3/T + 20.140\ln{T}$  & mol.dm$^{-3}$.atm$^{-1}$      \\ 
Kr\tnote{b}       & ln(H$_i$)= $-174.52 + 9101.7/T + 24.221\ln{T}$  & mol.dm$^{-3}$.atm$^{-1}$      \\ 
Xe\tnote{b}       & ln(H$_i$)= $-197.21 + 10521.0/T + 27.466\ln{T}$ & mol.dm$^{-3}$.atm$^{-1}$      \\ 
\hline
\end{tabular}
\label{tab:henrys_value}
\begin{tablenotes}
\item [a] \cite{kawazuishi_correlation_1987}.
\item [b] \cite{warneck_atmospheric_2012}.
\end{tablenotes}
\end{threeparttable}
\end{table*}

The molar volume of the species $i$ at infinite dilution, $v_{i,H_2O}^{\infty}$, is extrapolated from \cite{rumpf_experimental_1993,rumpf_solubility_1993} and \cite{garcia_density_2001} for CO$_2$ and NH$_3$. For the other species, $v_{i,H_2O}^{\infty}$ is set to 0. $R$ is the ideal gas constant. 

\noindent Following other studies in the same pressure-temperature range \citep{pazuki_prediction_2006,marounina_role_2018}, we compute the fugacity coefficient $\phi_i$ for each species using the Peng-Robinson equation of state (EOS) \citep{peng_new_1976} and the Van der Waals one-fluid mixing rule. 
The molar volume of the mixture $v$ is derived from the Peng-Robinson EOS: 
\begin{equation}
P = \cfrac{RT}{v-b} - \cfrac{a\alpha}{v^2 +2vb - b^2},
\label{eq:EOSPG}
\end{equation}

\noindent where 

\begin{equation}
\begin{split}  
&a = \cfrac{0.457235R^2T_c^2}{P_c},\\
&b = \cfrac{0.077796RT_c}{P_c},\\
&\alpha = (1+\kappa(1-\sqrt{\cfrac{T}{T_c}}))^2,\\
{\rm and}~&\kappa = 0.37464 + 1.54226\omega - 0.2699\omega^2.
\end{split}
\label{eq:parametersEOS}
\end{equation}

\noindent Here, $T_c$ is the critical temperature, $P_c$ is the critical pressure, and $\omega$ is the acentric factor of each volatile. Their values are displayed in Table \ref{tab:critical_values}.

The mixture parameters $A_{mix}$ and $B_{mix}$ required for the Peng-Robinson equation of state and the Van der Waals one-fluid mixing rule are computed using the following equations:

\begin{equation}
\begin{split}
&A_i=a_i\times\alpha_i,\\
&A_{ij} = (1-k_{ij})\sqrt{A_iA_j},\\
&A_{mix} = \sum_{i}^{N}\sum_{j}^{N} y_iy_jA_{ij},\\
{\rm and}~&B_{mix} = \sum_{i}^{N}y_ib_i,
\end{split}
\end{equation}

\noindent where $N$ is the total number of species. $\alpha_i$, $a_i$ and $b_i$ are computed for each species $i$ using Eq. \ref{eq:parametersEOS}. When available, binary interaction coefficients $k_{ij}$ are used to further account for the non-ideal interactions between the volatile species, following the approach described in \cite{gasem_modified_2001}. If not, they are set to 0. These binary interaction coefficients are determined empirically and taken from the literature (\cite{dhima_solubility_1999} for H$_2$O-CO$_2$, and \cite{vrabec_molecular_2009} for the other couples).
By incorporating these binary interaction coefficients, the model can more accurately capture the non-ideal behavior of the volatile mixtures in the atmosphere and ocean.

\begin{table}[htpb]
\centering
\caption{$T_c$, $P_c$, and $\omega$ values used in the model \citep{reid_properties_1987}.}
\begin{tabular}{@{}cccc@{}}
\hline
\hline
\smallskip
Molecule & $T_c$ & $P_c$ & $\omega$ \\
\hline
\smallskip
H$_2$O   & 647.3    & 220.6      & 0.3434 \\
CO$_2$   & 304.19   & 73.83      & 0.224  \\
NH$_3$   & 405.6    & 112.8      & 0.25    \\ 
O$_2$    & 154.4    & 50.5       & 0.022   \\ 
CO       & 132.9    & 35.0       & 0.066   \\ 
CH$_4$   & 190.4    & 46.0       & 0.011   \\ 
N$_2$    & 126.3    & 33.9       & 0.039   \\ 
Ar       & 150.8    & 48.7       & 0.001   \\ 
Kr       & 209.4    & 55.0       & 0.005   \\ 
Xe       & 289.7    & 58.4       & 0.008   \\ 
\hline
\end{tabular}
\label{tab:critical_values}
\end{table}

After having computed the molar volume $v$ from the Peng-Robinson equation of state (Eq. \ref{eq:EOSPG}), the fugacity coefficient $\phi_i$ of a species $i$ can be derived from Eq. \ref{eq:EOSPG}:

\begin{equation}
\begin{split}
    \ln(\phi_i(T,v,y)) &= \cfrac{b_i}{B_{mix}}(Z-1) - \ln[\cfrac{P}{RT}(v-B_{mix})] + \\
    &\cfrac{A_{mix}}{2\sqrt{2}RTB_{mix}}\ln(\cfrac{v - B_{mix}2.414}{v- B_{mix}0.414})(\delta_i - \cfrac{b_i}{B_{mix}})
\end{split}
\end{equation}

\noindent with, 

\begin{empheq}{align}
&Z = \cfrac{Pv}{RT},
&\delta_i = 2\frac{A_i}{A_{mix}}\sum_{j}y_j\sqrt{A_j}(1-k_{ij}).
\end{empheq}

The activity coefficients of H$_2$O, CO$_2$ and NH$_3$ are computed with the extended UNIQUAC model \citep{thomsen_modeling_1999}. The activity coefficient of CH$_4$ is taken from \cite{kvamme_small_2021}. For the other neutral species, we set $\gamma_i=1$.

\subsection{Chemical equilibrium}

In this model we couple the liquid-vapor equilibrium with the chemical equilibrium taking place within the aqueous H$_2$O-CO$_2$-NH$_3$ system:  

\begin{align}
        H_2O &\rightleftharpoons{} H^+ + OH^-, \\
        \label{eq:carbonate}
        CO_2 + H_2O &\rightleftharpoons{} HCO_3^- + H^+, \\
        \label{eq:ammonium}
        NH_3 + H_2O &\rightleftharpoons{} NH_4^+ + OH^-, \\
        \label{eq:bicarbonate}
        HCO_3^- &\rightleftharpoons{} CO_3^{2-} + H^+, \\
        \label{eq:carbamate}
        HCO_3^- + NH_3 &\rightleftharpoons{} NH_2COO^- + H_2O.    
\end{align}

The equations describing the chemical equilibrium within the H$_2$O-CO$_2$-NH$_3$ system are derived from the dissociation reactions of these species :

\begin{align}
    K_{H_2O} &= \cfrac{m_{OH^-}m_{H^+}}{x_{H_2O}} \; \cfrac{\gamma_{OH^-}\gamma_{H^+}}{\gamma_{H_2O}}, \\
    K_{CO_2} &= \cfrac{m_{HCO_3^-}m_{H^+}}{x_{H_2O}m_{CO_2}} \; \cfrac{\gamma_{HCO_3^-}\gamma_{H^+}}{\gamma_{H_2O}\gamma_{CO_2}}, \\
    K_{NH_3} &= \cfrac{m_{NH_4^{+}}m_{OH^-}}{x_{H_2O}m_{NH_3}} \; \cfrac{\gamma_{NH_4^+}\gamma_{OH^-}}{\gamma_{H_2O}\gamma_{NH_3}}, \\
    K_{HCO_3^-} &= \cfrac{m_{CO_3^{2-}}m_{H^+}}{m_{HCO_3^-}} \; \cfrac{\gamma_{CO_3^{2-}}\gamma_{H^+}}{\gamma_{HCO_3^{-}}}, \\
    K_{NH_2COO^-} &= \cfrac{m_{NH_2COO^-}x_{H_2O}}{m_{NH_3}m_{HCO_3^-}} \; \cfrac{\gamma_{NH_2COO^-}\gamma_{H_2O}}{\gamma_{NH_3}\gamma_{HCO_3^-}},
\end{align}

\noindent where $K_i(T)$ is the dissociation constant of each reaction taken from \cite{kawazuishi_correlation_1987}, $m_i$ (in mol/kg) is the molality of the aqueous species, $x_{H_2O}$ the mole fraction of water, and  $\gamma_i$ the activity coefficient of the species $i$.

The activity coefficients involved in the chemical equilibrium equations are computed using the extended UNIQUAC model, first introduced in \cite{thomsen_modeling_1999} and the parameters of \cite{darde_modeling_2010}. This model calculates the activity coefficient using combinatorial, residual, and electrostatic terms, with the latter based on the extended Debye-Hückel law. It is well established for studying this specific chemical equilibrium \citep{thomsen_modeling_1999,thomsen_modeling_2005,darde_modeling_2010}.
Finally, to retrieve the whole set of variables, the mass and charge balance equations are derived as follows: 

\begin{align}
&x_{CO_{{2}{tot}}} = x_{CO_{2}} + x_{HCO_{3}^-} + x_{CO_{3}^{2-}} + x_{NH_{2}COO^{-}}, \\
&x_{NH_{{3}{tot}}} = x_{NH_{3}} + x_{NH_{4}^{+}} + x_{NH_{2}COO^{-}}, \\
&x_{H^{+}} + x_{NH_{4}^{+}} = x_{OH^{-}} + x_{HCO_{3}^{-}} + 2x_{CO_{3}^{2-}} + x_{NH_{2}COO^{-}},
\label{eq:balance}
\end{align}

\noindent where $x_{CO_{{2}{tot}}}$ and $x_{NH_{{3}{tot}}}$ are the total mole fractions of CO$_2$ and NH$_3$ incorporated into the liquid phase. We do not account for the charges brought by Na$^+$ and Cl$^-$ in Eq. \ref{eq:balance} as their concentrations in Europa's ocean, particularly in its primordial state, remain poorly constrained. Moreover, the concentration ranges reported in the literature vary depending on the composition of the rocky seafloor and the water-to-rock ratio, and these geochemical factors are beyond the scope of this paper.

These equations ensure the conservation of mass and charge within the H$_2$O-CO$_2$-NH$_3$ system. The resulting system of equations, coupling the liquid-vapor equilibrium and the chemical equilibrium, is solved using the Levenberg-Marquart minimization method. This equation-solving process is highly sensitive to the initial guesses, so the model has been adjusted and bench-marked against UNIQUAC model results and experimental data from the literature \citep{van_krevelen_composition_1949,goppert_vaporliquid_1988,bieling_evolutionary_1989,darde_modeling_2010}.

Experimental data \citep{van_krevelen_composition_1949,goppert_vaporliquid_1988,bieling_evolutionary_1989,darde_modeling_2010} and models (e.g \cite{castillorogez_contribution_2022}) show that when CO$_2$ and NH$_3$ are present together in water, they can react to form ammonium carbamate (NH$_2$COO$^-$), as described in \cref{eq:carbamate}. \cite{van_krevelen_composition_1949} pointed out that below a certain ratio of total dissolved CO$_2$ to NH$_3$ m$_{CO_2}$/m$_{NH_3}$ $\sim$ [0.4--0.5]), ammonia can trap CO$_2$ in the liquid phase, preventing it from entering the gaseous phase. In this scenario, the dissolved CO$_2$ is converted into carbonate and bicarbonate ions (HCO$_3^-$, CO$_3^{2-}$) \cref{eq:carbonate,eq:bicarbonate}, which then react with ammonium (NH$_4^+$) \cref{eq:ammonium} to form ammonium carbamate (NH$_2$COO$^-$) \cref{eq:carbamate}. However, above this threshold, a thick CO$_2$ atmosphere can be in equilibrium with the water ocean.

\section{Results} \label{sec:results}

In the following, we investigate two different sets of initial conditions for our model. The first set explores the scenario where Europa's hydrosphere originates from the accretion of building blocks with a comet-like composition. The second set of initial conditions assumes that Europa's primordial hydrosphere formed in equilibrium with a thick, CO$_2$-rich atmosphere.

\subsection{Hydrosphere formation from comet-like building blocks}
\label{sec3.1}

\begin{table}[htpb]
\centering
  \begin{threeparttable}[b]
\caption{Volatile distributions from cometary building blocks adopted for the primordial outgassed atmosphere (expressed in mole fractions)}
\begin{tabular}{@{}ccc@{}}
\hline
\hline
\smallskip
Molecule & 67P\tnote{1}      & CO$_2$-rich comet\tnote{2}\\
\hline
\smallskip
H$_2$O   & $9.18\times10^{-1}$    & $7.28\times10^{-1}$ \\
CO$_2$   & $4.31\times10^{-2}$   & $2.18\times10^{-1}$ \\
NH$_3$   & $6.15\times10^{-3}$  & $4.87\times10^{-3}$ \\
O$_2$    & $2.84\times10^{-2}$   & $2.25\times10^{-2}$ \\
CO       & $2.84\times10^{-2}$   & $2.25\times10^{-2}$ \\
CH$_4$   & $3.12\times10^{-3}$  & $2.47\times10^{-3}$ \\
N$_2$    & $8.17\times10^{-4}$  & $6.48\times10^{-4}$ \\ 
Ar       & $5.32\times10^{-6}$  & $4.22\times10^{-6}$ \\
Kr       & $4.50\times10^{-7}$  & $3.57\times10^{-7}$ \\
Xe       & $2.20\times10^{-7}$  & $1.75\times10^{-7}$ \\ 
\hline
\label{tab:compositions}
\end{tabular}
     \begin{tablenotes}
       \item [1] \cite{rubin_elemental_2019}.
       \item [2] \cite{bockelee-morvan_composition_2017}.
     \end{tablenotes}
  \end{threeparttable}
\end{table}

Europa could have been partly formed from volatile-rich solids presenting a cometary-like composition \citep{canup_formation_2002,ronnet_pebble_2017,mousis_early_2023}. The potential distribution of volatiles brought to Europa's hydrosphere can be assumed to span a range of comet compositions, as discussed in \cite{bockelee-morvan_composition_2017}.

The model described above is used to explore how the initial volatile dispatch, particularly the total dissolved CO$_2$/NH$_3$ ratio, can influence the partitioning of volatiles between the atmosphere and the ocean of early Europa and  similar icy bodies.

We have applied the model to two different volatile distributions in the ice phase of comets, shown in Table \ref{tab:compositions}. Case 1a is derived from data from Comet 67P/Churyumov-Gerasimenko \citep{rubin_elemental_2019}. Case 1b is an end-member composition, with the same gas distribution as 67P, except CO$_2$ which is set to the maximum fraction observed in comets \citep{bockelee-morvan_composition_2017}.
The calculation is done at a temperature of 300K, which is a plausible surface temperature reached after accretion \citep{bierson_explaining_2020}. The initial total pressure is approximated to be $P_{\rm tot}$=$P^{\rm sat}_{\rm H_2O}$, with water being the main constituent of the atmosphere right after accretion.

\begin{figure*}[htpb]
\centering
\includegraphics[width=17cm]{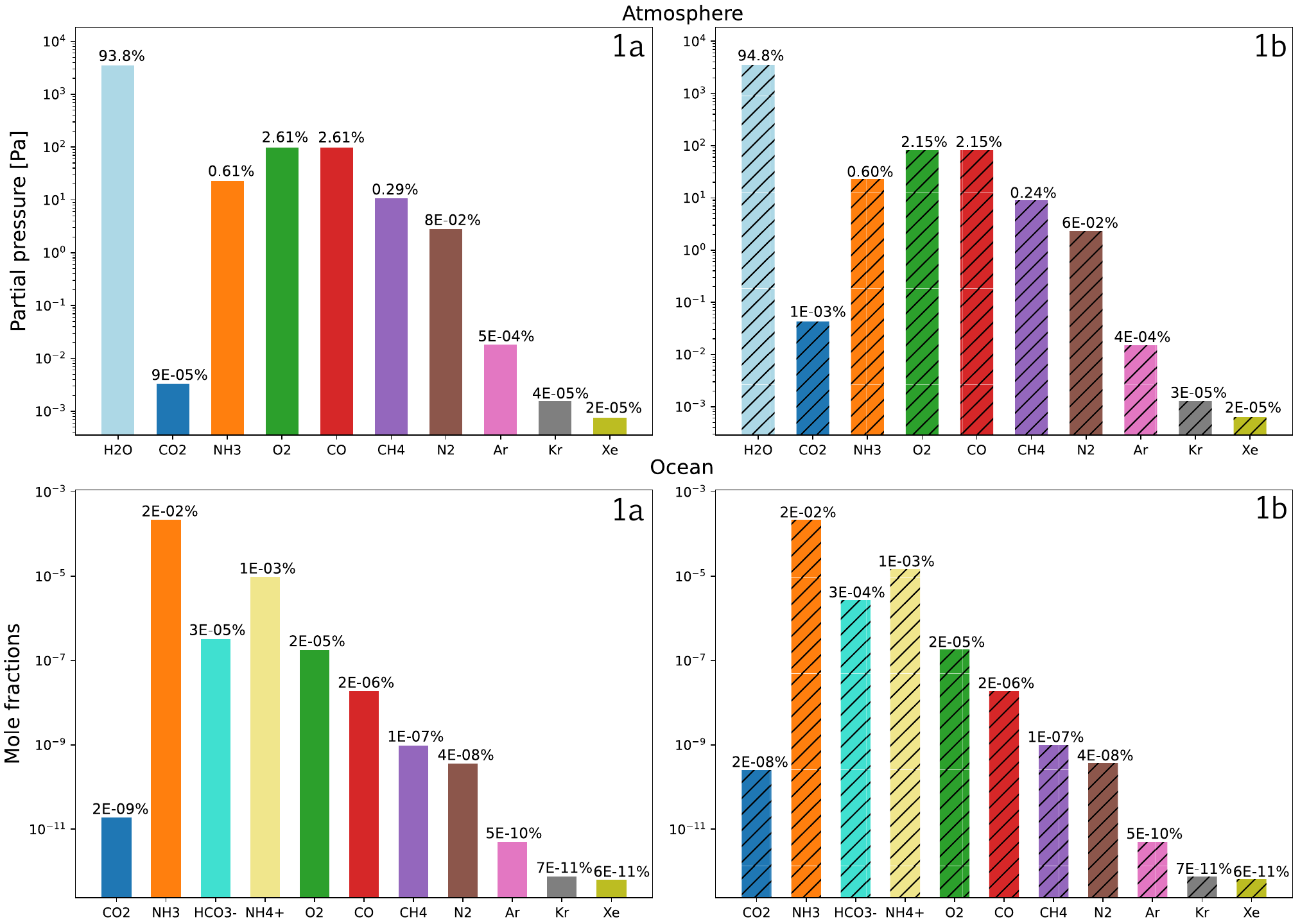}
\caption{Volatile abundances at equilibrium in the primordial atmosphere and ocean of Europa at $T$~=~300K. Two initial distributions are considered: Case 1a (solid), based on the composition of Comet 67P, and Case 1b (hatched), derived from a high CO$_2$ end-member composition. Final atmospheric fractions are expressed as partial pressures (Pa), while dissolved volatile fractions are presented as mole fractions. The percentage share of each species in each phase is displayed above each vertical bar. The figure highlights only the two most abundant ions, HCO$_3^-$ and NH$_4^+$.} 
\label{fig:cometary_compo}
\end{figure*}

The resulting distribution of species at equilibrium between the two phases is shown in Fig. \ref{fig:cometary_compo}. For the case in which Europa's hydrosphere formed from cometary ice, a significant fraction of ammonia is initially introduced into the primordial atmosphere.  
While carbon dioxide is initially the second most abundant component of the atmosphere after water vapor, its solubility in water is lower than that of ammonia. Consequently, a larger fraction of ammonia initially dissolves in the ocean compared to carbon dioxide when exposed to the atmosphere. 
Figure \ref{fig:cometary_compo} shows that the consequence of this small total dissolved CO$_2$/NH$_3$ ratio is the scarce amount of CO$_2$ left in the atmosphere and dissolved in the ocean once chemical equilibrium is reached. The pH at the interface of the ocean at equilibrium is alkaline in both cases (pH $\sim 10.5$). Indeed, at such high pH, CO$_2$ is mostly retained as bicarbonate and carbonate ions.

\smallskip

\begin{figure*}[htpb]
\centering
\includegraphics[width=11cm]{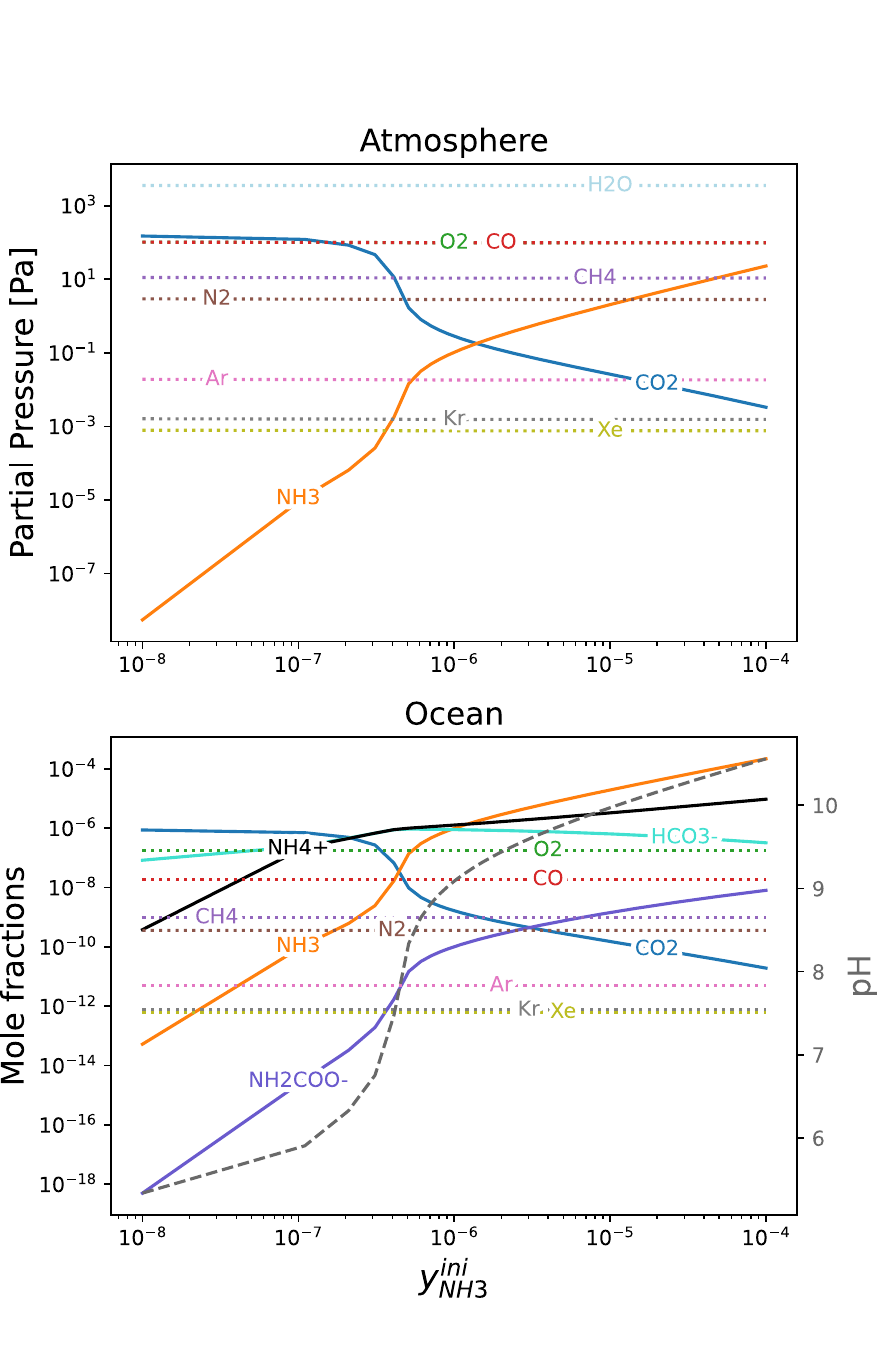}
\caption{Evolution at $T$ = 300 K of the final distribution of species in the atmosphere and ocean at equilibrium as a function of $y_{NH_3}^{ini}$, the initial mole fraction of NH$_3$ incorporated into the atmosphere. The initial mole fractions of all other species remain constant. Only the three most abundant ions--HCO$_3^-$, NH$_4^+$, and NH$_2$COO$^-$--are displayed in this figure. The dashed grey line corresponds to the pH of the ocean calculated as a function of $y_{NH_3}^{ini}$.}
\label{fig:NH3 variation}
\end{figure*}

To further highlight the influence of the initial atmospheric CO$_2$/NH$_3$ ratio on the distribution of volatiles in the primordial atmosphere, Fig. \ref{fig:NH3 variation} represents the sensitivity of the distribution of the molecules at $T$ = 300K to the variation of the fraction of NH$_3$ initially incorporated into the atmospheric interface. The fractions of CO$_2$ and other species initially incorporated into the gas phase are constant. y$_{NH_3}$ is taken in the range of $10^{-8}$ to $6.15\times10^{-3}$ (Table \ref{tab:compositions}).

Figure \ref{fig:NH3 variation} highlights the existence of a threshold on the initially accreted CO$_2$/NH$_3$ ratio at which most of the CO$_2$ is transformed, both in the atmosphere and in the ocean. Indeed we compute that, in our case, the share of NH$_3$ prevails over CO$_2$ in atmosphere when initial $y_{NH_3}^{ini} > 10^{-4}\times y_{CO_2}$. As the initial fraction of NH$_3$ increases, so does the fraction of N-bearing ions, especially NH$_2$COO$^-$, which is formed from carbonate ions. In fact, the chemical reaction that forms NH$_2$COO$^-$ is a thermodynamically favored reaction (K$_{NH_2COO^-}$ $\sim$2 whereas K$_{HCO_3^-}$ $\sim$ 10$^{-11}$ at 300K). As the fraction of NH$_3$ increases, the concentration of HCO$_3^-$ remains relatively constant, indicating that CO$_2$ is increasingly transformed to maintain chemical equilibrium. The figure shows that CO$_2$ and NH$_3$ cannot coexist as the primary atmospheric compounds simultaneously, as previously found by \cite{marounina_role_2018} in their study of Titan's primordial ocean. Their relative abundances exhibit an inverse relationship, where increasing the initial NH$_3$ abundance in the atmosphere leads to a corresponding decrease in CO$_2$.
Furthermore, we can emphasize the relationship between this trend and the ocean's pH. As predicted by equations \ref{eq:carbonate}--\ref{eq:carbamate}) and visible in Fig. \ref{fig:NH3 variation}, a higher CO$_2$/NH$_3$ ratio acidifies the ocean, while an increase of ammonia favors the formation of carbonates and a basic pH. Hence the initially accreted CO$_2$/NH$_3$ ratio is a pivotal factor in the determination of the ocean's pH as well.

\begin{figure*}[htpb]
\centering
\includegraphics[width=11cm]{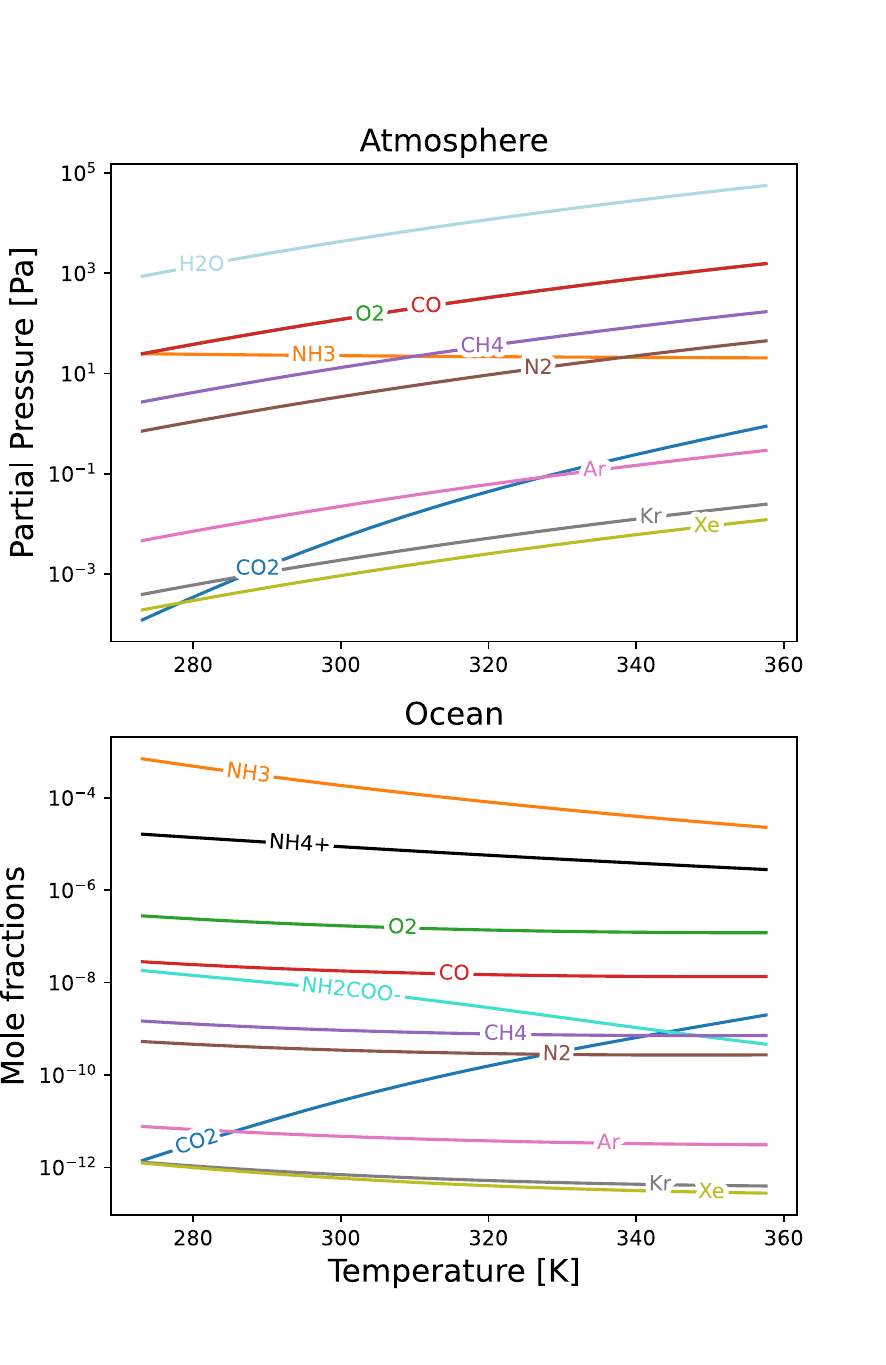}
\caption{Evolution of final partial pressures and mole fractions in the ocean at equilibrium as a function of temperature. Not all dissolved ions in the ocean are displayed in the figure, even though they are present in the mixture.} 
\label{fig:Tvariation}
\end{figure*}

Figure \ref{fig:Tvariation} illustrates the impact of temperature on the chemical equilibrium shown in Case 1a, investigated over a range from 273.15 K to 358 K. The upper limit of 358 K represents a plausible temperature that can be attained on Europa after accretion. We compute it using the approach outlined in \cite{schubert_internal_1981} and \cite{grasset_cooling_1996}. This calculation considers Europa's kinetic energy from accretion, with a fraction $h$ of this energy—specifically, $h$ = 0.1 —converted into heat. 

The initial atmospheric pressure at the interface is approximated by the saturation vapor pressure of water. Figure \ref{fig:Tvariation} shows that, as the initial pressure increases with temperature \citep{bridgeman_vapor_1964}, the partial pressures of the various species also increase by about an order of magnitude in pressure between 273.15 K and 358 K. However, due to temperature-dependent variations in Henry's constants, the mole fractions of the dissolved species, including the noble gases (tracers of the moon's evolution), show a slight decrease with temperature, less than an order of magnitude.  However, CO$_2$ and NH$_3$ behave differently from the other species because of the chemical reactions in which they are involved. In fact, with increasing temperature, CO$_2$ is less likely to be trapped by the formation of NH$_2$COO$^-$, because this chemical reaction is less thermodynamically favored with increasing temperature \citep{kawazuishi_correlation_1987}. Thus, at higher temperatures, a larger fraction of CO$_2$ remains in the atmosphere (more than two orders of magnitude in partial pressure between 273.15 K and 357.6 K). Consequently, the partial pressure of NH$_3$ decreases slightly, and the mole fraction of dissolved NH$_3$ lessens by an order of magnitude with increasing temperature. 

\subsection{Equilibrium with a thick and CO$_2$--dominated atmosphere}

\begin{table}[htpb]
\centering
\caption{Adopted volatile distributions in the primordial atmospheric reservoir, assuming equilibration with a CO$_2$--rich atmosphere (expressed in mole fractions)}

\begin{tabular}{@{}ccc@{}}
\hline
\hline
\smallskip
Molecule & P$_{CO_2}=1$ bar   & P$_{CO_2}=10$ bar \\
\hline
\smallskip
H$_2$O   & $3.42\times10^{-2}$  & $3.36\times10^{-3}$ \\
CO$_2$   & $9.63\times10^{-1}$  & $9.96\times10^{-1}$ \\
NH$_3$   & $2.23\times10^{-4}$  & $2.40\times10^{-5}$ \\
O$_2$    & $1.11\times10^{-3}$  & $1.10\times10^{-4}$ \\
CO       & $1.11\times10^{-3}$  & $1.10\times10^{-4}$ \\
CH$_4$   & $1.12\times10^{-4}$  & $1.20\times10^{-5}$ \\
N$_2$    & $3.00\times10^{-5}$  & $3.00\times10^{-6}$ \\ 
Ar       & $1.98\times10^{-7}$  & $2.05\times10^{-8}$ \\
Kr       & $1.67\times10^{-8}$  & $1.73\times10^{-9}$ \\
Xe       & $8.22\times10^{-9}$  & $8.50\times10^{-10}$ \\ 
\hline
\end{tabular}
\label{tab:compositions_pCO2high}
\end{table}

\begin{figure*}[htpb]
\centering
\includegraphics[width=17cm]{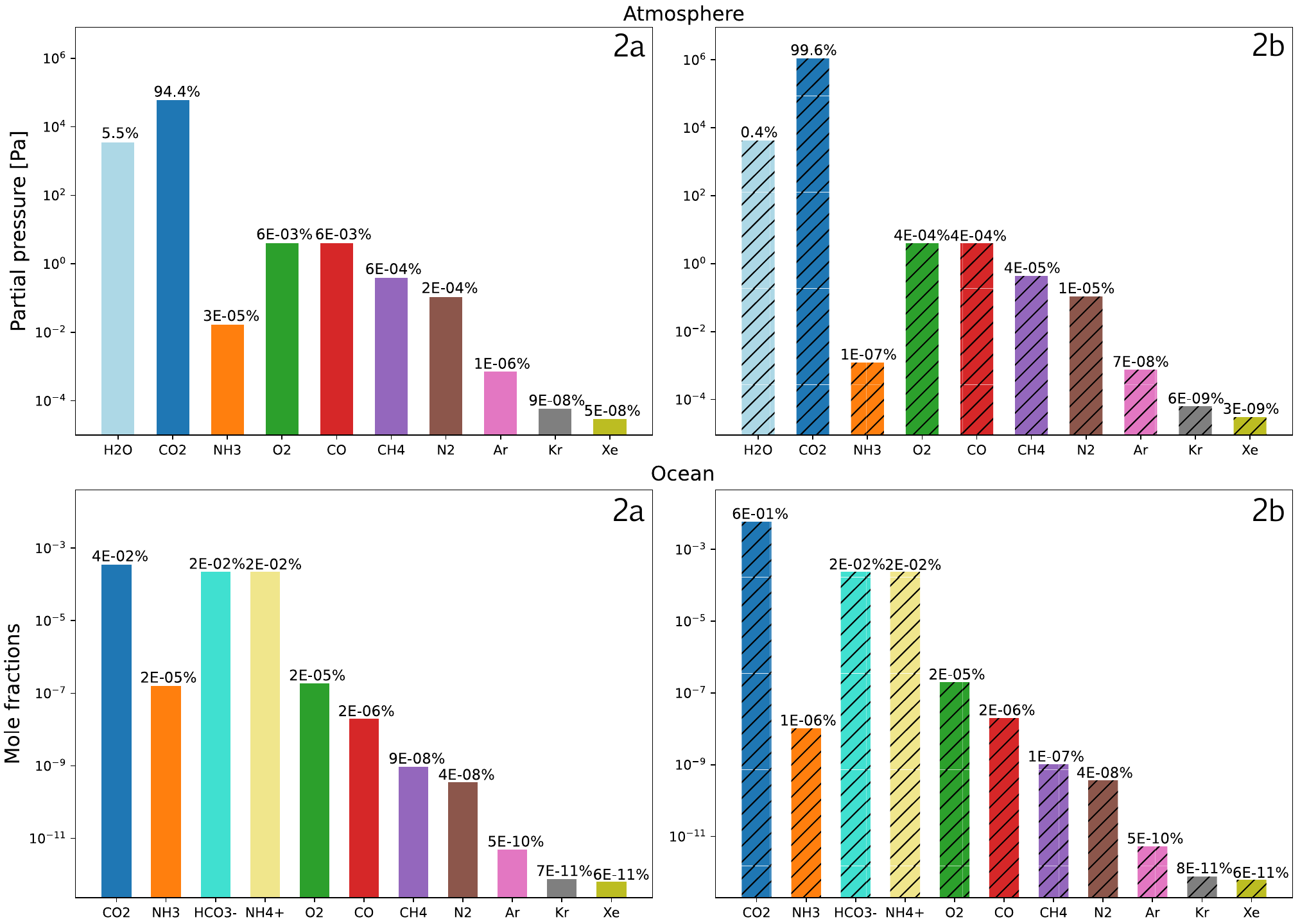}
\caption{Volatile abundances at equilibrium in the primordial atmosphere and ocean of Europa at $T$ = 300 K. Two initial distributions are considered: one with \(P_{CO_2}\) initially set to 1 bar (Case 2a, solid) and another with \(P_{CO_2}\) initially set to 10 bar (Case 2b, hatched). Atmospheric fractions are represented as partial pressures (Pa), while dissolved volatile fractions are shown as mole fractions. The figure displays only the two most abundant ions, HCO$_3^-$ and NH$_4^+$.}
\label{fig:Compo_highCO2}
\end{figure*}

Here, it is assumed that Europa's primordial hydrosphere equilibrated with a thick and CO$_2$--rich atmosphere, based on the idea that the ocean could originate from a metamorphic origin \citep{melwani_daswani_metamorphic_2021}.
As previously shown in Fig. \ref{fig:NH3 variation}, the dominance of the partial pressure of carbon dioxide over that of ammonia, and vice versa, is determined by the accreted CO$_2$/NH$_3$ ratio in the ocean. Fig. \ref{fig:Compo_highCO2} illustrates the equilibrium state of the primordial hydrosphere for two initial atmospheric conditions: one with the partial pressure of CO$_2$ set to 1 bar (comprising 96.3\% of the atmosphere) (Case 2a) and the other set to 10 bar (comprising 99.6\% of the atmosphere) (Case 2b). The initial partial pressure of water is set to the saturation vapor pressure of water, and the partial pressures of other species are calculated as fractions of this saturation vapor pressure, following the fractions provided by \cite{rubin_elemental_2019}. The corresponding initial volatile fractions are listed in Table \ref{tab:compositions_pCO2high}.

Starting with a high CO$_2$ fraction in the primordial atmosphere results in a higher fraction of total accreted CO$_2$ compared to NH$_3$. Consequently, such a configuration maintains a CO$_2$-rich primordial atmosphere at equilibrium. However, contrary to what is shown in Sec. \ref{sec3.1}, with the total dissolved CO$_2$/NH$_3$ ratio favoring CO$_2$, almost all NH$_3$ is transformed in both scenarios. The speciation of NH$_3$ in NH$_4^+$ and NH$_2$COO$^-$ leads to a significantly reduced residual fraction of free NH$_3$ in both phases at equilibrium compared to the initial conditions in the primordial atmosphere.

Additionally, the higher the initial CO$_2$ partial pressure, the less CO$_2$ is removed from the atmosphere due to chemical equilibrium. When the initial $P_{CO_2}$ is 1 bar, 40\% of the initial CO$_2$ is sequestered in ionic forms in the ocean. In contrast, this percentage drops to nearly 0\% when the initial $P_{CO_2}$ is 10 bar. Unlike the cases studied in Table \ref{tab:compositions}, the pH at the ocean-atmosphere interface resulting from this equilibrium is acidic, with pH values of 6.1 for Case 2a and 4.9 for Case 2b.

\section{Discussion}
\label{sec:discussion}

The composition of Europa's current subsurface ocean remains poorly constrained \citep{becker_exploring_2024}. However, surface observations in recent years have provided insights into the variety of species present within the ocean \citep{ligier_vltsinfoni_2016,hibbitts_color_2019,trumbo_distribution_2023,villanueva_endogenous_2023}. The concentration ranges of these dissolved species are uncertain, and models depend on only partially understood physical variables such as pH, salinity, and rock composition. Furthermore, due to the potential role of radiolytic oxidation of organics on Europa's surface, assessing the extent of aqueous alteration of compounds like CO$_2$ is challenging \citep{zolotov_aqueous_2012} which in turn complicates the estimation of the amount of carbonates present in the ocean. Although they were targeted, carbonate content could not be estimated by \cite{trumbo_distribution_2023} and \citep{villanueva_endogenous_2023}.

Our study uncovers the implications of several formation scenarios on the state of the primordial hydrosphere, particularly regarding the distribution of ionic compounds resulting from the speciation of CO$_2$ and NH$_3$. Current knowledge of Europa's subsurface ocean does not allow us to favor one end-member scenario over another regarding its primordial hydrosphere composition. However, \cite{zolotov_chemical_2009} modeled examples of Europa's ocean composition based on the ocean's pH and redox state, demonstrating that the pH is highly dependent on the types of rocks (e.g., chondritic, basaltic) exposed to water. 

It is also important to emphasize that the type and quantity of dissolved gases influence the ocean's pH (see Fig. \ref{fig:NH3 variation}), as predicted by thermodynamic principles. Our computations indicate that when CO$_2$ is primarily sequestered as carbonates due to high NH$_3$ concentrations, the ocean maintains a basic pH at equilibrium (Case 1a,b). In contrast, when in equilibrium with a thick CO$_2$--rich atmosphere (Case 2a,b), the ocean exhibits a more acidic pH. 

As highlighted by geochemical models, the water--to--rock ratio and the reactivity of non-volatile compounds are crucial in determining the ocean's pH. \cite{zolotov_aqueous_2012} noted that when the igneous rocks of the silicate mantle are exposed to water, the types of minerals subject to dissolution—and thus involved in the ocean's chemistry—depend on whether the ocean is in acidic or alkaline conditions. In an alkaline environment, CO$_2$ can be converted into carbonates and bicarbonates (Case 1a,b), which may eventually precipitate. However, as demonstrated by \cite{zolotov_aqueous_2012}, carbonate concentrations could be influenced by the presence of ions such as Ca$^{2+}$, allowing for other components to impact the CO$_2$-NH$_3$ chemical equilibrium via:

\begin{align}
        2HCO_3^- + Ca^{2+} &\rightleftharpoons{} CaCO_3 + H_2O +CO_2, \\
        \label{eq:preci_bicarbonate}
        CO_3^{2-} + Ca^{2+} &\rightleftharpoons{} CaCO_3.
\end{align}

\noindent On the other hand, salinity can play a key role in determining the concentration of precipitated solutes, which in turn affects the pH. \cite{castillorogez_contribution_2022} emphasized that the presence of carbonates and other ions resulting from the equilibrium between CO$_2$, NH$_3$, and rocks can also influence the salinity and conductivity of the ocean. These parameters may have been impacted by changes in the thickness of the icy shell over time as well \citep{melosh_temperature_2004}. Although extreme fluctuations are unlikely due to the presence of buffering agents \citep{johnson_insights_2019}, the current state of knowledge regarding the ocean's composition does not permit a reliable assessment of Europa's oceanic pH \citep{becker_exploring_2024}. Without this information, we are unable to constrain the CO$_2$/NH$_3$ ratio.

Future measurements are needed to better assess the state of Europa's subsurface ocean. As highlighted by \cite{villanueva_endogenous_2023} and \cite{trumbo_distribution_2023}, observations of surface features and spectral analyses can test the endogenic origin of species detected through remote sensing. If future space missions identify plumes on Europa's surface, in situ measurements could provide direct data on materials originating from the ocean, similar to what Cassini accomplished with the Enceladus plumes \citep{postberg_sodium_2009,postberg_salt-water_2011,waite_cassini_2017,glein_geochemistry_2018,hansen_composition_2020}.

Specifically, in situ data from Europa Clipper's SUDA and MASPEX instruments \citep{waite_maspex-europa_2024}, along with JUICE's PEP instrument, will offer valuable insights into ions such as HCO$_3^-$ and CO$_3^{2-}$, which are pertinent to our study, as well as neutral species present in the ocean \citep{fohn_description_2021,becker_exploring_2024}. These measurements could be complemented by remote sensing analyses from Clipper's Europa--UVS instrument to gather additional information on volatiles. By integrating data on CO$_2$ from MASPEX and PEP and carbonates from SUDA, we could constrain the primordial CO$_2$/NH$_3$ inventory thanks to our model. Thus, modeling the relationship between the current and primordial states of Europa's ocean will enable more accurate estimates of the relative abundances of specific species (especially CO$_2$ and NH$_3$), helping to differentiate between competing end-member scenarios based on the results of this study.

\section{Summary and conclusions}\label{sec:summary}

Our work provides an initial assessment of the distribution of primordial volatiles in Europa's early hydrosphere, considering the liquid-vapor equilibrium at the atmosphere-ocean interface and the chemical equilibrium within the ocean. To do so, we assumed that the atmospheric reservoir outgassed from the interior due to vigorous accretional and radiogenic heating, reaching equilibrium with an underlying liquid ocean. We show how the initial NH$_3$ proportion in Europa's building blocks could influence the primordial atmospheric CO$_2$ distribution. In case 1, the initial volatile atmospheric distribution was based on cometary abundances. We demonstrate that at 300K, the primordial hydrosphere at equilibrium is CO$_2$--depleted, despite a significant fraction being delivered to the early atmosphere. By varying the initial NH$_3$ fraction, we observe that a CO$_2$--rich atmosphere could be sustained above a certain CO$_2$/NH$_3$ threshold. However in case 2, the initial primordial atmosphere's composition is CO$_2$-rich. We showed that starting from such a configuration, the primordial atmosphere retains a significant CO$_2$ proportion at equilibrium, while NH$_3$ is transformed in the liquid phase. Overall we highlight, studying those two cases, the significance of the initially accreted CO$_2$/NH$_3$ ratio on their partial pressure distribution at equilibrium in the primordial atmosphere. The proportion of CO$_2$ in the atmosphere is correlated to the value of this ratio, which can lead to a depletion of multiple orders of magnitude in partial pressure when it is below the threshold.
This conclusion relates to the amount of NH$_3$ that could have initially been incorporated into Europa's hydrosphere, which may have helped keep the ocean liquid due to its antifreeze properties \citep{neveu_aqueous_2017}. However, the required NH$_3$ abundance must be balanced against the CO$_2$ quantity needed to match present-day abundances \citep{trumbo_distribution_2023,villanueva_endogenous_2023}. A significant concentration of dissolved NH$_3$ (greater than 5\%) is required to effectively lower the freezing point of the ocean \citep{tobie_titans_2012}. Should future measurements indicate a substantial fraction of free CO$_2$ dissolved in the ocean, this will impose an upper limit on NH$_3$ concentration based on the results of this study. Therefore, if the NH$_3$ concentration falls below the threshold necessary for a significant antifreeze effect, the anti-freeze contribution for maintaining the ocean in a liquid state may need to be reconsidered.

In the presence of a significant amount of NH$_3$, the likely consequence, depending on the counter ions available, is that a significant amount of these carbonates will precipitate, accentuating the suppression of CO$_2$. The work of \cite{glein_carbonate_2020} on Enceladus showed that the carbonate content of the ocean can be strongly influenced by the balance between the ocean and  seafloor alteration minerals. \cite{melwani_daswani_metamorphic_2021} also explored how rock devolatilization and aqueous alteration, taking into account carbonate precipitation, can lead to a thick CO$_2$ primordial atmosphere for Europa. However, NH$_3$ has not been included in these models, and we have shown that its presence can shift the resulting equilibrium. It is therefore crucial to further couple this model with a geochemical model to explore how exchanges between the rocky bottom of Europa's ocean may affect its composition. In addition, as a consequence of the extraction of elements from the silicate mantle, additional species may be introduced into the ocean and affect the latter's equilibrium as well \citep{zolotov_composition_2001,vance_geophysical_2016}, including the nature and amount of antifreeze such as NH$_3$ \citep{neveu_aqueous_2017}. Although we have only considered the chemical reactions involving CO$_2$ and NH$_3$ in this model, more chemistry should be considered in the case of hydrothermal processes. Indeed, \cite{seewald_experimental_2006} showed that reduced carbon compounds could be formed under these conditions, with CO$_2$, CH$_3$OH and CO as major or trace components. 

One should note that this study's model focuses on the primordial ocean-atmosphere interaction.
Moreover, the atmosphere considered in this model is isothermal and without escape. Such assumptions limit the accuracy of the model, since volatile species such as CO and CH$_4$ are present. Atmospheric escape would profoundly affect the hydrosphere equilibrium, depleting the atmosphere of volatiles. In addition, the greenhouse effect that could be produced by the gas has not been included. However, our model provides an upper limit on the partial pressures reached in the atmosphere without escape, for the most volatile species released by the building blocks after accretion.
\smallskip

Additionally, the formation of clathrate hydrates should be considered under the following conditions: 

\begin{enumerate}
\item If an additional source of volatiles raises the concentration in the ocean above the solubility limit, leading to bubble formation.
\item If the pressure of the gaseous mixture in the bubbles exceeds the dissociation pressure of the corresponding clathrate hydrate.
\end{enumerate}

\noindent Clathrate hydrates could serve as another volatile reservoir, influencing the composition of the primordial atmosphere over time \citep{prieto-ballesteros_evaluation_2005, zolotov_chemical_2009, mousis_abundances_2013, bouquet_role_2019}. If the density of clathrate hydrates is lower than that of the ocean, they can float to the surface, forming a crustal layer that affects the exchange between the liquid and gaseous phases \citep{prieto-ballesteros_evaluation_2005, marounina_role_2018, kalousova_insulating_2020}. The buoyancy of clathrates can be influenced by the ocean's salinity and composition \citep{prieto-ballesteros_evaluation_2005}, highlighting the importance of coupling a self-consistent, time-dependent model of clathrate hydrate formation with the one presented in this paper.

Our work provides a first assessment of the distribution of primordial volatiles in Europa's primitive hydrosphere. The \textit{Europa Clipper} mission will be able to measure the current composition of Europa's hydrosphere. By combining these measurements with future modeling, it will be possible to derive the composition of the primordial hydrosphere and compare it with the predictions of our model, providing insight into the conditions under which Europa formed.

\begin{acknowledgments}
O.M. and A.B. acknowledge support from CNES. The project leading to this publication has received funding from the Excellence Initiative of Aix-Marseille Université – A*Midex, a French “Investissements d’Avenir programme” AMX-21-IET-018. This research holds as part of the project FACOM (ANR-22-CE49-0005-01 ACT) and has benefited from a funding provided by l’Agence Nationale de la Recherche (ANR) under the Generic Call for Proposals 2022. C.R.G.'s work was supported by NASA's Preparatory Science Investigations for Europa (PSIE) program.
\end{acknowledgments}

\vspace{5mm}

\bibliography{Ammonia_Europa.bib}{}
\bibliographystyle{aasjournal}

\end{document}